\documentclass[%
preprint,
superscriptaddress,
showpacs,
amsmath,amssymb,
prc
floatfix]%
{revtex4-1}

\usepackage{color}
\definecolor{mygray}{gray}{.9}
\definecolor{mypink}{rgb}{.99,.91,.95}
\definecolor{mycyan}{cmyk}{.3,0,0,0}
\usepackage{graphicx}
\usepackage{dcolumn}
\usepackage{bm}

\begin{document}







\title{High-K isomer and the rotational properties in the odd-Z neutron-rich nucleus $^{163}$Eu}

\author{Xiao-Tao He}%
\email{hext@nuaa.edu.cn}
 \affiliation{College of Material Science and Technology, Nanjing University of Aeronautics and Astronautics, Nanjing 210016, China}
\author{Ze-Long Chen}%
 \affiliation{College of Material Science and Technology, Nanjing University of Aeronautics and Astronautics, Nanjing 210016, China}




\begin{abstract}
The newly observed isomer and ground-state band in the odd-Z neutron-rich rare-earth nucleus $^{163}$Eu are investigated by using the cranked shell model (CSM) with pairing treated by the particle-number conserving (PNC) method. This is the first time detailed theoretical investigations are performed of the observed $964(1)$ keV isomer and ground-state rotational band in $^{163}$Eu. The experimental data are reproduced very well by the theoretical results. The configuration of the $964(1)$ keV isomer is assigned as the three-particle state $\frac{13}{2}^{-}(\nu\frac{7}{2}^{+}[633]\otimes\nu\frac{1}{2}^{-}[521]\otimes\pi\frac{5}{2}^{+}[413]$). More low-lying multi-particle states are predicted in $^{163}$Eu. Due to its significant effect on the nuclear mean field, the high-order $\varepsilon_{6}$ deformation plays an important role in the energy and configuration assignment of the multi-particle states. Compared to its neighboring even-even nuclei $^{162}$Sm and $^{164}$Gd, there is a $10\%\sim15\%$ increase of $J^{(1)}$ of the one-particle ground-state band in $^{163}$Eu. This is explained by the pairing reduction due to the blocking of the nucleon on the proton $\pi\frac{5}{2}^{+}$[413] orbital in $^{163}$Eu.
\end{abstract}



\maketitle


\section{Introduction}
In recent years, due to the extensive application of radioactive detecting devices in nuclear experiments, it has become possible to study the structure of neutron-rich nuclei far from the $\beta$-stability line. The in-beam spectroscopy of neutron-rich nuclei in the rare earth region has been successfully investigated~\cite{YokoyamaR2017_PRC95,PatelZ2017_PRC96,IdeguchiE2016_PRC94,SoederstroemP2016_PLB762,PatelZ2016_PLB753,PatelZ2016_EWoC123,WatanabeH2016_PLB760,PatelZ2014_PRL113,SimpsonG2009_PRC80,UrbanW2009_PRC80}. These studies revealed detailed information of the nuclear structure in exotic nuclei. With the increase of neutron number, changes in the shell structure are expected. Recently,  for example, Hartley et al. presented evidence of the existence of a deformed sub-shell gap at $N=98$~\cite{HartleyD2018_PRL120}, contrary to the previous understanding that such a gap occurred at $N=100$. Odd-mass nuclei can provide additional information for the Nilsson configuration assignment of the rotational band. However, most of the existing studies concentrate on the properties of even-even nuclei. Experimental data for odd-mass neutron-rich rare-earth nuclei are rare due to the experimental difficulties.

A level scheme based on isomer depopulation in odd-Z neutron-rich nucleus $^{163}$Eu was presented recently by Yokoyama \textit{et al}.~\cite{YokoyamaR2017_PRC95}, and independently by Patel \textit{et al}.~\cite{PatelZ2017_PRC96}. The newly observed $\gamma-$ray spectrum was assigned to the ground-state band based on the proton $\pi\frac{5}{2}^{+}$[413] state. The $964(1)$ keV isomer was interpreted as a three-particle state, while its configuration is still an open issue. So far, there is no detailed theoretical calculations of these experimental investigations.

In the present work, the newly observed isomer and rotational band built on the ground-state of odd-Z neutron rich nucleus $^{163}$Eu are investigated by the cranked shell model, with pairing treated by the particle-number conserving method. This is the first time that a spectroscopical study including both the isomer and ground-state band in $^{163}$Eu is performed theoretically. To discuss the effect of the unpaired odd nucleon on the rotational properties of $^{163}$Eu, calculations of the neighboring even-even nuclei $^{162}$Sm and $^{164}$Gd are carried out as well.

\section{Theoretical framework}
The cranked shell model Hamiltonian with pairing correlation is $H_\mathrm{CSM}= H_{\rm SP}-\omega J_x + H_\mathrm{P}$. $H_{\rm SP}=\sum_{\xi}(h_\mathrm{Nil})_{\xi}$ is the single-particle Hamiltonian where $h_{\mathrm{Nil}}$ is the Nilsson Hamiltonian and $\xi$ ($\eta$) the eigen-state of the single-particle Hamiltonian $h_{\xi(\eta)}$, and $\bar{\xi}$ ($\bar{\eta}$) its time-reversed state. $-\omega J_x$ is the Coriolis interaction with the cranking frequency $\omega$ about the $x$ axis (perpendicular to the nuclear symmetry z axis). The pairing $H_{\text{P}}=H_\mathrm{P}(0)+H_\mathrm{P}(2)$ includes monopole $H_{\text{P}}(0)=-G_{0}\sum_{\xi \eta }a_{\xi }^{\dagger}a_{\overline{\xi }}^{\dagger }a_{\overline{\eta }}a_{\eta }$ and quadrupole $H_{\text{P}}(2)=-G_{2}\sum_{\xi \eta } q_{2}(\xi) q_{2}(\eta)a_{\xi}^{\dagger } a_{\overline{\xi}}^{\dagger} a_{\overline{\eta}}a_{\eta}$ pairing correlations, where $q_{2}(\xi) = \sqrt{{16\pi}/{5}}\langle \xi |r^{2}Y_{20} | \xi \rangle$ is the diagonal element of the stretched quadrupole operator.

In the PNC method, the cranked shell model Hamiltonian $H_\mathrm{CSM}$ is diagonalized in the cranked many-particle configuration (CMPC) space without particle quasi-particle transformation. Thus the particle-number is conserved and the Pauli blocking effect is taken into account exactly. The CMPC space is a Fock space. Therefore, the Hamiltonian $H_\mathrm{CSM}$ can be diagonalized in a comparatively small space to obtain sufficiently accurate solutions of the ground and low-lying states.

The eigenstate of $H_{\textrm{CSM}}$ is $| \psi \rangle = \sum_{i} C_i | i \rangle$, with the configuration $| i \rangle$ defined by the occupation of the cranked single-particle orbitals by the real particles. The solution $| \psi \rangle$ can always be obtained even for a pair-broken state, and provides a reliable way of assigning the configuration of a multi-particle state. $n_\mu=\sum_{i}|C_{i}|^{2}P_{i\mu}$ is the occupation probability of the cranked orbital $|\mu\rangle$, where $P_{i\mu}=1$ if $|\mu\rangle$ is occupied in $|i\rangle$, and $P_{i\mu}=0$ otherwise.
The particle number is $N=\sum_\mu n_\mu$. The kinematic moment of inertia in the state $|\psi\rangle$ is given by $J^{(1)} = \frac{1}{\omega} \langle \psi | J_x | \psi \rangle$, where the angular momentum alignment $\left\langle \psi \right| J_{x}
 \left| \psi \right\rangle=\sum_{i}\left|C_{i}\right| ^{2}
   \left\langle i\right| J_{x}\left| i\right\rangle
 + 2\sum_{i<j}C_{i}^{\ast }C_{j}
   \left\langle i\right| J_{x}\left| j\right\rangle\ $. Details of the PNC-CSM method can be found in Refs.\cite{WuC1989_PRC39, ZengJ1994_PRC50, XinX2000_PRC62}.

\section{Results and discussions}{\label{Sec:results}}

In the present work, the Nilsson states are calculated for the valence single-particle states for proton $N = 0-5$ and neutron $N = 0-6$ major shells. The Nilsson parameters $(\kappa,\mu)$ are taken from the Lund systematics~\cite{NilssonS1969_NPA131}. The deformation parameters $\varepsilon_{2}$, $\varepsilon_{4}$ and $\varepsilon_{6}$ are taken from the M\"oller's table~\cite{MoellerP1995_ADaNDT59}, except for $\varepsilon_{4}$ in $^{163}$Eu. Instead of $\varepsilon_{4}=0$, $\varepsilon_{4}=-0.02$ is taken to reproduce better the experimental excitation energy and moment of inertia of $^{163}$Eu.

The effective pairing strengths $G_{0}$ and $G_{2}$ can be determined, in principle, by the odd-even differences of the nuclear binding energies in principle. For the neutron-rich rare-earth nuclei, their values are determined by the odd-even differences of moments of inertia. The effective pairing strengths are connected with the dimension of the truncated CMPC space. In the present calculations, the CMPC spaces are constructed for the proton $N = 3, 4, 5$ and neutron $N = 4, 5, 6$ shells. The dimension of the CMPC space is about 700, and the corresponding effective pairing strengths are $G_{0} = 0.20$ MeV and $G_{2} = 0.02$ MeV for both neutrons and protons. Effective monopole and quadrupole pairing forces with similar strengths are used in the Projected Shell Model~\cite{SunY2016_PS91}. The stability of the PNC-CSM calculations with the variation of the dimension of the CMPC space was investigated in Refs.~\cite{ZengJ1983_NPA405,LiuS2002_PRC66,ZhangZ2012_PRC85}.

\begin{figure}
\begin{center}
\includegraphics[width=9cm]{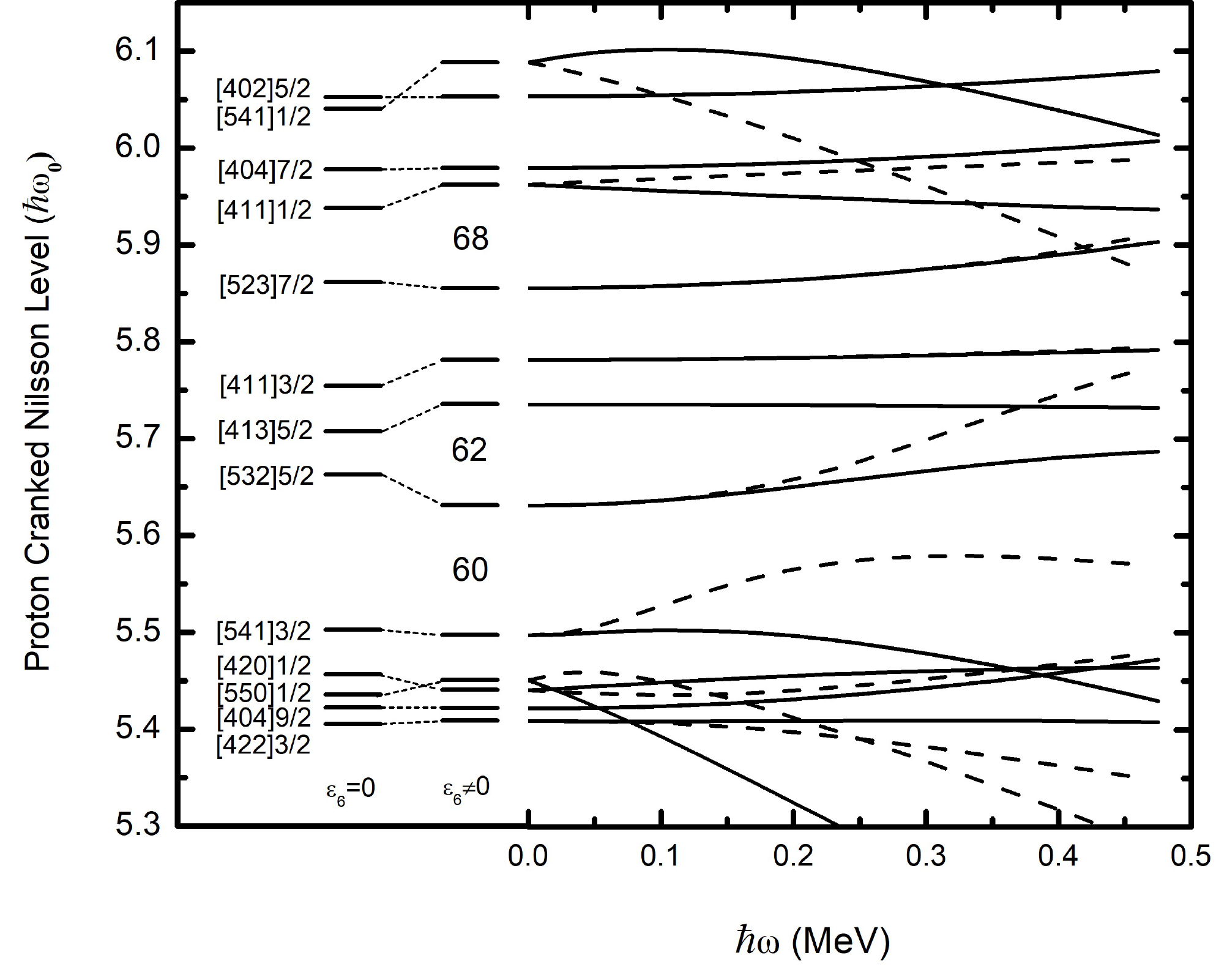}
\caption{\label{fig:Fig1} Cranked proton Nilsson levels near the Fermi surface of $^{163}$Eu with signature $\alpha=+1/2$ (solid) and $\alpha=-1/2$ (dash). $\varepsilon_{2}=0.275$, $\varepsilon_{4}=-0.02$ and $\varepsilon_{6}$=0.042. }
\end{center}
\end{figure}

The cranked proton single particle level structure near the Fermi surface of $^{163}${Eu} is very similar to the neighboring even-even nuclei~\cite{HeX2018_PRC98}, and is presented in Fig.~\ref{fig:Fig1}. The signature $\alpha=+1/2(-1/2)$ levels are denoted by solid (dashed) lines. Results with and without high-order deformation $\varepsilon_{6}$ are compared at rotational frequency $\hbar\omega=0$. A deformed energy gap at $Z = 62$ arises in calculations with non-zero $\varepsilon_{6}$, which leads to a significant effect on the energy and configuration assignment of the multi-particle states, especially for the newly discovered $964(1)$ keV isomer. This will be discussed in detail later. In addition, compared to the results with zero $\varepsilon_{6}$, the deformed energy gap at $Z=60$ is reduced and the one at $Z=68$ enlarged. Our results in Fig.~\ref{fig:Fig1} show that the inclusion of the $\varepsilon_{6}$ deformation can change the order of some single particle levels, resulting in appearance of new sub-shell gaps. To reflect the shell changes in the exotic mass region, an early attempt was made by empirically adjusting the Nilsson parameters from their standard values, see for example Refs.~\cite{ZhangZ2012_PRC85,SunY2000_PRC62,ZhangJ1998_PRC58}.

\begin{figure}
\begin{center}
\includegraphics[width=13.5cm]{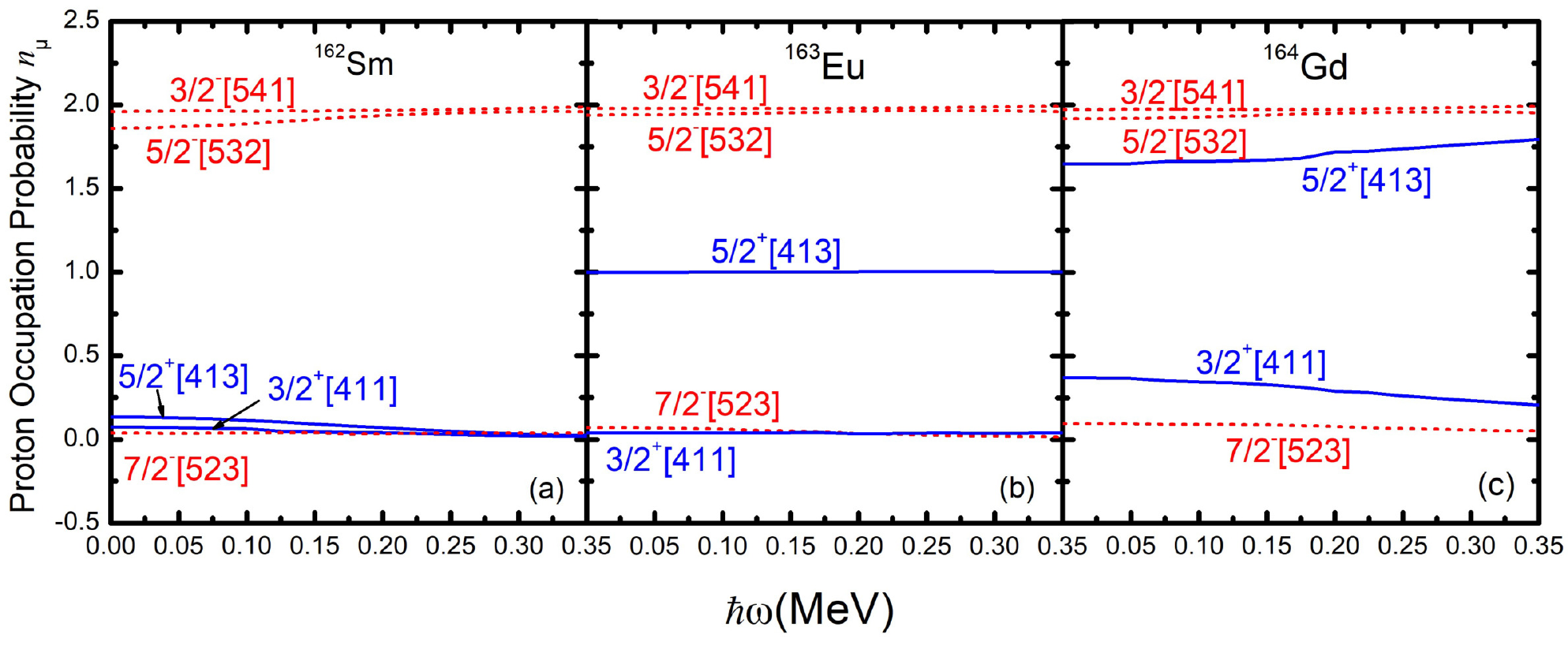}
\caption{\label{fig:Fig2}  (Color online) 
Occupation probabilities $n_\mu$ of each cranked proton orbital $\mu$ (including both $\alpha = \pm1/2$) near the Fermi surface of $^{162}$Sm, $^{163}$Eu and $^{164}$Gd for the ground-state bands. The solid blue (short dash red) line denotes positive (negative) parity orbital. Fully occupied $n_{\mu}\approx2$ and empty $n_{\mu}\approx0$ orbitals are not labelled.}
\end{center}
\end{figure}

The configuration of each multi-particle state is explicitly assigned through the occupation probability $n_{\mu}$ of each cranked Nilsson orbital $\mu$. The proton orbital occupation probabilities $n_{\mu}$ versus rotational frequency $\hbar\omega$ of the ground-state band in $^{162}$Sm, $^{163}$Eu and $^{164}$Gd are shown in Fig.~\ref{fig:Fig2}. As shown, the proton orbital $\pi\frac{5}{2}^{+}[413]$ is blocked $(n_{\mu}\approx1)$ in $^{163}$Eu while it is either almost fully occupied $(n_{\mu}\approx2)$ or empty $(n_\mu\approx0)$ in even-even nuclei  $^{162}$Sm and $^{164}$Gd. Therefore, the configuration of the ground-state band in $^{163}$Eu is assigned as $\pi\frac{5}{2}^{+}[413]$. Configurations of the other multi-particle states, listed in Table.~\ref{tab:163Eu}, are assigned similarly.

\begin{table*}
\centering
\caption{Low-lying multi-particle states in $^{163}$Eu predicted by the PNC-CSM calculations.}
\begin{ruledtabular}
\begin{tabular}{ccccc}
 $K^\pi$ & Configuration  & $E_x$(keV)($\varepsilon_{6}\neq0$) &$E_x$(keV)($\varepsilon_{6}=0$)& $E_{x}^{exp}$(keV) \\
 \hline
    $\frac{5}{2}^{+}$   &$\pi\frac{5}{2}^{+}[413]$                     &0      &0      &0\\
    $\frac{3}{2}^{+}$   &$\pi\frac{3}{2}^{+}[411]$                     &312.1  &273.7  & \\
    $\frac{5}{2}^{-}$   &$\pi\frac{5}{2}^{-}[532]$                     &778.9  &400.9  & \\
    $\frac{7}{2}^{-}$   &$\pi\frac{7}{2}^{-}[523]$                     &851.0  &1050.2 & \\
    $\frac{1}{2}^{+}$   &$\pi\frac{1}{2}^{+}[411]$                     &1617.9 &1630.5 & \\
    $\frac{7}{2}^{+}$   &$\pi\frac{7}{2}^{+}[404]$                     &1738.0 &       & \\
    $\frac{1}{2}^{+}$   &$\pi\frac{1}{2}^{+}[420]$                     &       &1858.5 & \\
    $\frac{3}{2}^{-}$   &$\pi\frac{3}{2}^{-}[541]$                     &2016.5 &1842.2 & \\
    $\frac{1}{2}^{-}$   &$\pi\frac{1}{2}^{-}[541]$                     &2523.6 &       & \\
    $\frac{13}{2}^{-}$  &$\nu\frac{7}{2}^{+}[633]\otimes\nu\frac{1}{2}^{-}[521]\otimes\pi\frac{5}{2}^{+}[413]$&1001.3&1057.0 &964(1)  \\
    $\frac{11}{2}^{-}$  &$\nu\frac{7}{2}^{+}[633]\otimes\nu\frac{1}{2}^{-}[521]\otimes\pi\frac{3}{2}^{+}[411]$&1313.4&1330.7 &     \\
    $\frac{13}{2}^{-}$  &$\pi\frac{5}{2}^{+}[413]\otimes\pi\frac{5}{2}^{-}[532]\otimes\pi\frac{3}{2}^{+}[411]$&1509.9&1134.9 &      \\
    $\frac{13}{2}^{+}$  &$\nu\frac{7}{2}^{+}[633]\otimes\nu\frac{1}{2}^{-}[521]\otimes\pi\frac{5}{2}^{-}[532]$&1780.2&1457.9 &     \\
    $\frac{17}{2}^{-}$  &$\nu\frac{7}{2}^{+}[633]\otimes\nu\frac{5}{2}^{-}[512]\otimes\pi\frac{5}{2}^{+}[413]$&1839.2&1499.3 &     \\
    $\frac{17}{2}^{+}$  &$\pi\frac{5}{2}^{+}[413]\otimes\pi\frac{5}{2}^{-}[532]\otimes\pi\frac{7}{2}^{-}[523]$&2024.5&1872.5 &     \\
    $\frac{17}{2}^{+}$  &$\nu\frac{7}{2}^{+}[633]\otimes\nu\frac{5}{2}^{-}[512]\otimes\pi\frac{5}{2}^{-}[532]$&2618.1&1900.2 &     \\
    \end{tabular}
\end{ruledtabular}
\label{tab:163Eu}
\end{table*}

The 964(1) keV isomer in $^{163}$Eu was observed recently by Yokoyama \textit{et al}.~\cite{YokoyamaR2017_PRC95} and independently by Patel \textit{et al}.~\cite{PatelZ2017_PRC96}. The spin and parity of this isomer is in both works given as $\frac{13}{2}^{-}$. However, its configuration is disputed. It was interpreted as the coupling of the $K^{\pi}=4^{-} (\nu^{2}\frac{1}{2}^{-}\frac{7}{2}^{+})$ neutron excitation and the $\pi\frac{5}{2}^{+}$[413] odd proton by the deformed Hartree-Fock model with angular momentum projection~\cite{YokoyamaR2017_PRC95}, while it was referred to as the three-proton excitation state with a configuration $\pi\frac{5}{2}^{+}$[413]$\otimes\pi\frac{5}{2}^{-}$[532]$\otimes\pi\frac{3}{2}^{+}$[411] in the Nilsson-BCS calculations~\cite{PatelZ2017_PRC96}.

The low-lying multi-particle states of $^{163}$Eu predicted by the PNC-CSM calculations are listed in Table~\ref{tab:163Eu}. A significant influence of the $\varepsilon_{6}$ deformation is demonstrated by the energy and configuration assignments of the multi-particle states. As shown in Table~\ref{tab:163Eu}, the lowest three-particle excitation state of $^{163}$Eu is the $\nu\frac{7}{2}^{+}[633]\otimes\nu\frac{1}{2}^{-}[521]\otimes\pi\frac{5}{2}^{+}[413]$ configuration state. Its energy is 1001.3 keV and 1057.0 keV as given by calculations with non-zero and zero $\varepsilon_{6}$, respectively. Both values reproduce well the experimental measurement. As for the second three-particle excitation state, if we do not consider the $\varepsilon_{6}$ effect, it is a $\pi\frac{5}{2}^{+}[413]\otimes\pi\frac{5}{2}^{-}[532]\otimes\pi\frac{3}{2}^{+}[411]$ state, which is predicted as the configuration of the 964(1) keV isomer by the Nilsson-BCS calculations. Its energy is 1134.9 keV in PNC-CSM calculations, which is very close to the experimental data as well. If we take into account the effect of $\varepsilon_{6}$ deformation, the second three-particle state is a $\nu\frac{7}{2}^{+}[633]\otimes\nu\frac{1}{2}^{-}[521]\otimes\pi\frac{3}{2}^{+}[411]$ configuration state. The three-proton state with a configuration $\pi\frac{5}{2}^{+}[413]\otimes\pi\frac{5}{2}^{-}[532]\otimes\pi\frac{3}{2}^{+}[411]$ becomes the third three-particle state with a much higher energy of 1509.9 keV, which can not reproduce well the experimental data. Thus the 964(1) keV $\frac{13}{2}^{-}$ isomer can be assigned as the $\nu\frac{7}{2}^{+}[633]\otimes\nu\frac{1}{2}^{-}[521]\otimes\pi\frac{5}{2}^{+}[413]$ configuration with high confidence. The rise of the three-proton state in non-zero $\varepsilon_{6}$ calculations results from the enlarged energy gap at $Z=62$ (see Fig.~\ref{fig:Fig1}). 

In general, compared to the calculations with zero $\varepsilon_{6}$, the state energies are higher in the non-zero $\varepsilon_{6}$ calculations for most of the three-particle states (see Table.~\ref{tab:163Eu}). Normally, the calculations with more degrees of freedom give lower energy. In the present calculations, with non-zero $\varepsilon_{6}$, the ground-state is lower in energy by about 5 MeV for protons and 15 MeV for neutrons. The eigen-energies of the excited states are lower as well. The higher energies of the multi-particle states in non-zero $\varepsilon_{6}$ calculations are mainly caused by the enlarged proton $Z=62$ and neutron $N=102$ deformed energy gaps when the $\varepsilon_{6}$ deformation is included. Several investigations have already given evidence that the high-order $\varepsilon_{6}$ deformation is important for the structure of the neutron-rich rare-earth nuclei~\cite{PatelZ2014_PRL113, PatelZ2016_PLB753, HeX2018_PRC98}.


\begin{figure}
\begin{center}
\includegraphics[width=9.2cm]{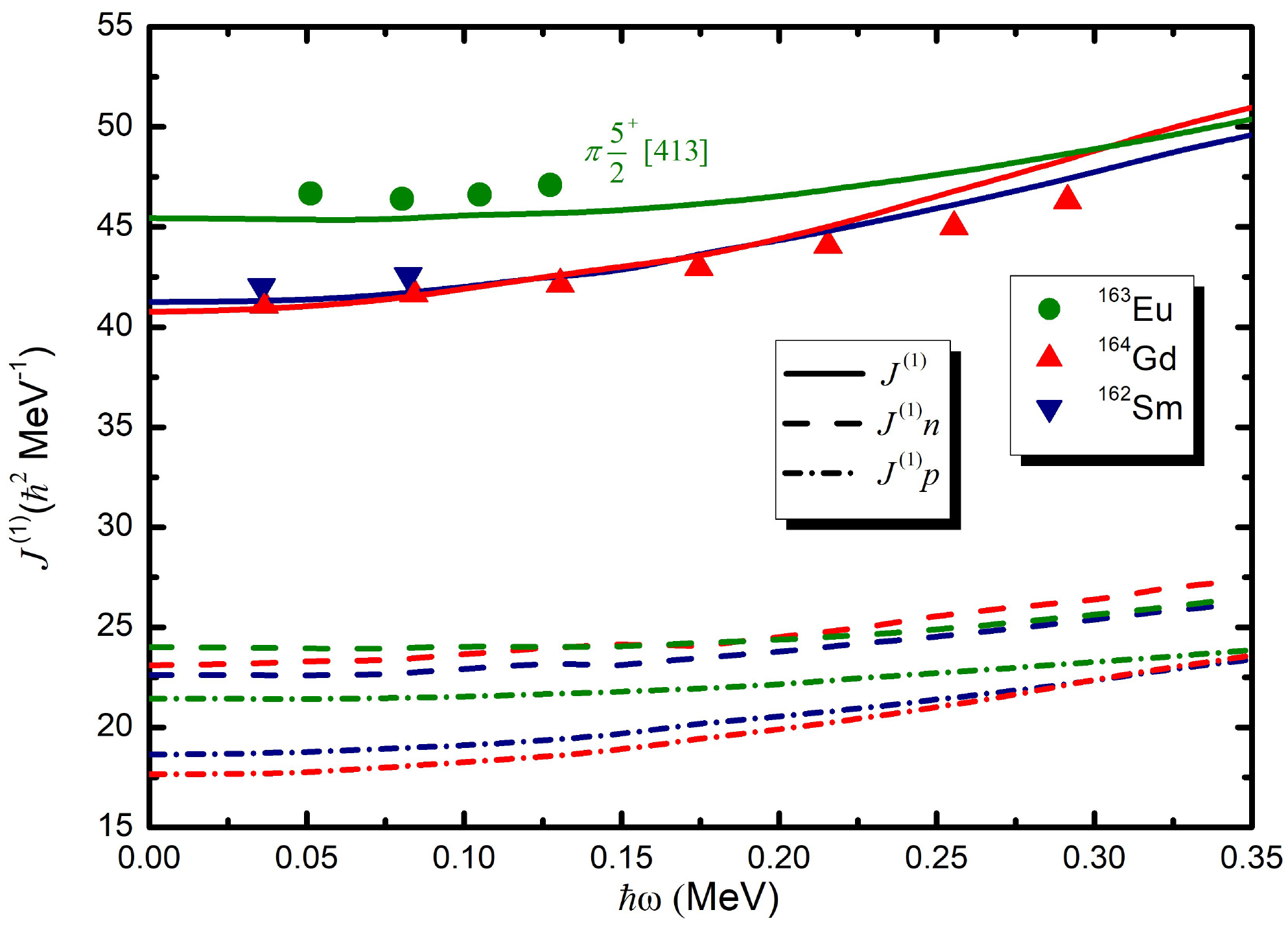}
\caption{\label{fig:Fig3}  Comparison of the theoretical kinematic moment of inertia $J^{(1)}$ for the ground-state bands in $^{162}$Sm (navy), $^{163}$Eu (olive) and $^{164}$Gd (red) with the experimental data. Theoretical total moments of inertia $J^{(1)}$ are denoted by the solid lines, the contribution from protons (neutrons) $J^{(1)}_{p}$ ($J^{(1)}_{n}$) are denoted by dashed dot (dashed) lines and the experimental data are denoted by symbols. }
\end{center}
\end{figure}

Fig.~\ref{fig:Fig3} shows the experimental and calculated kinematic moment of inertia $J^{(1)}$ for the ground-state bands in $^{162}$Sm, $^{163}$Eu and $^{164}$Gd. No significant signature splitting is found in these bands, and only the calculated moments of inertia for favored signature bands are shown. It is seen that the experimental data are reproduced quite well by the PNC calculations. Compared to the neighboring even-even nuclei $^{162}$Sm and $^{164}$Gd, a $10\%\sim15\%$ increase of $J^{(1)}$ can be seen for the one-particle ground-state band in $^{163}$Eu. This can be explained by the pairing reduction due to blocking. Calculations without pairing, which are not shown here, result in moments of inertia $J^{(1)}$ that are almost the same for these three bands. $J^{(1)}$ of the ground-state bands in $^{162}$Sm and $^{164}$Gd are very similar, although there are two more protons in $^{164}$Gd than in $^{162}$Sm. Additional information is given by the contributions of protons $J^{(1)}_{p}$ (dashed dot lines) and neutrons $J^{(1)}_{n}$ (dashed lines). $J^{(1)}_{n}$ for all three bands is almost the same, while $J^{(1)}_{p}$ for the one-particle ground-state band in $^{163}$Eu is larger than in $^{162}$Sm and $^{164}$Gd by $\sim18\%$ in the low frequency range. The $10\%\sim15\%$ increase in $J^{(1)}$ for $^{163}$Eu comes from the contribution of protons, namely, from the Pauli blocking effect of the proton $\pi\frac{5}{2}^{+}[413]$ orbital.

\section{Summary}{\label{Sec:summary}}

The recently observed isomer and ground-state band in odd-Z neutron-rich rare-earth nucleus $^{163}$Eu are investigated by using the cranked shell model, with pairing treated by a particle-number conserving method. The experimental energy of the isomer and the kinematic moment of inertia $J^{(1)}$ of the ground-state band are reproduced very well by the theoretical calculations. This is the first time a detailed spectroscopical investigation of the observed $964(1)$ keV isomer and ground-state of $^{163}$Eu is performed theoretically. To investigate the pairing and Pauli blocking effects, the rotational bands in the neighboring even-even nuclei $^{162}$Sm and $^{164}$Gd were calculated as well.

The configuration of the isomer is assigned as the two-neutron excitation of $K^{\pi}=4^{-} (\nu^{2}\frac{1}{2}^{-}\frac{7}{2}^{+})$ coupled with the odd-proton $\pi\frac{5}{2}^{+}$[413] state, i.e.,  $\frac{13}{2}^{-}(\nu\frac{7}{2}^{+}[633]\otimes\nu\frac{1}{2}^{-}[521]\otimes\pi\frac{5}{2}^{+}[413]$) configuration state, which is consistent with the assignment of the deformed Hartree-Fock model with angular momentum projection. The high-order $\varepsilon_{6}$ deformation plays an important role in the configuration assignment due to its effect on the nuclear mean field. More low-lying multi-particle states in $^{163}$Eu are predicted.

Compared to the neighboring even-even nuclei $^{162}$Sm and $^{164}$Gd, a $10\%\sim15\%$ increase of $J^{(1)}$ is found in the ground-state band of $^{163}$Eu. Detailed theoretical investigations show that the increase of $J^{(1)}$ comes from the contribution of protons. It can be explained by the pairing reduction due to the blocking of the odd-proton on the $\pi\frac{5}{2}^{+}$[413] orbital in $^{163}$Eu.

\begin{acknowledgements}
This work has been supported by the National Natural Science Foundation of China under grant No. 11775112 and the Priority Academic Program Development of Jiangsu Higher Education Institutions.
\end{acknowledgements}

\bibliographystyle{apsrev4-1}
\bibliography{../../../../References/ReferencesXT}

\end{document}